\documentclass[aps,preprint,showpacs,superscriptaddress,groupedaddress]{revtex4}  % for review and submission
\usepackage{graphicx}  % needed for figures
\usepackage{dcolumn}   % needed for some tables
\usepackage{bm}        % for math
\usepackage{amssymb}   % for math
\usepackage{amsmath}
\usepackage[font=small,labelfont=bf]{caption}
\newcommand{\be}{\begin{equation}}
\newcommand{\ee}{\end{equation}}

\begin{document}
\title{Primordial black holes, zero-point energy and the CMB: The cosmic connection}
\author{S. Zeynizadeh}\email{zeynizadeh@physics.sharif.ir}\affiliation{Department of Physics, Sharif University of Technology
P.O. Box 11365-9161, Tehran, Iran}\affiliation{School of physics, Institute for Research in Fundamental Sciences (IPM)
P.O. Box 19395-5531, Tehran, Iran}
\author{M. Nouri-Zonoz }\email{nouri@khayam.ut.ac.ir}\affiliation{Department of Physics, University of Tehran,
North Karegar Ave., P.O. Box 14395-547, Tehran, Iran.}
\date{\today}

\begin{abstract}
We propose a possible resolution to the cosmological constant problem through a scenario in which the universe is composed of three components: matter, radiation (CMB) and vacuum energy such that vacuum energy is not constant and is decaying into the matter component. Matter in this scenario consists of baryonic matter and primordial black holes (PBHs) as the dark matter. Local equilibrium condition between PBHs and CMB confines the mass and the radius of PBHs. The mechanism accounting for the decaying process is nothing but the absorption  of vacuum energy modes by the PBHs up to a wavelength of the order of their radius. Acting as a natural cut-off on the wavelength of vacuum energy modes, this leads to the observed value for the vacuum energy density.
\end{abstract}

\maketitle
%%%%%%%%%%%%%%%%%%%%%%%%%%%%%%%%%%%%%%%%%%%%%%%%%%%%%%%%%%
\section{Introduction}
One of the most fundamental problems in theoretical physics is the so called {\it cosmological constant problem}, which is formulated at the interface of cosmology and quantum field theory \cite{Weinberg:1989cp}. An interesting  approach towards a possible resolution of this problem is the introduction of a decaying cosmological {\it term} instead of a constant  one. This term, identified with vacuum energy density in Friedmann equations, is taken not to be a constant and decays into other components of the energy-momentum content of the universe namely dark matter or radiation \cite{Freese:1987dd, Overduin:1998zv}. None of the models based on decaying vacuum energy, presented as yet, are totally satisfactory either due to the lack of a consistent mechanism or because of their inconsistency with the observational data \cite{Overduin:1998zv} and that is why it is generally being discussed from phenomenological point of view \cite{Opher:2004}.

In the present essay we are going to introduce a consistent mechanism for the above mentioned decay by choosing a physically motivated cut-off frequency for the vacuum energy modes of an effective scalar field. In our scenario it is assumed that the energy-momentum content of the universe consists of three components, matter, radiation and vacuum energy of the above mentioned effective scalar field. Material part is composed of baryonic matter as the visible sector and primordial black holes (PBHs) as the dark sector \cite{Frampton:2010sw, Blais, Khlop} and the radiation is composed of cosmic microwave background (CMB) photons. Vacuum energy is given by the ground state energy of the effective scalar field modes integrated over all wavelengths from Hubble length as IR cut-off to Planck length as UV cut-off \cite{Martin, Paddy}. Planck length as the UV cut-off gives rise to a vacuum energy density which is about 121 orders of magnitude larger than the observed value. Therefore
one needs a physically motivated UV cut-off to arrive at the observed value for the vacuum energy density. In our scenario the mechanism  inducing such a cut-off is based on the following three assumptions:

1-There are PBHs in the universe distributed homogeneously and isotropically at the large scale.

2-Mass of these black holes is determined by their local thermal equilibrium (LTE) with the CMB.

3-Radius of these PBHs is a natural cut-off wavelength for the vacuum energy density, i.e  only 
those modes of vacuum  survive  whose wavelengths are larger than  the radius of the PBHs and the 
shorter wavelength modes are eaten up by the PBHs.

Obviously by calling them assumptions, each one of them could be questioned on different grounds including 
observational and theoretical constraints, but none of them pose a contradiction with the well stablished facts in the standard cosmology.
%%%%%%%%%%%%%%%%%%%%%%%%%%%%%%%%%%%%
\section{Order of magnitude calculation}
To examine the feasibility of the above proposed mechanism, an order of magnitude calculations for the vacuum energy density could be carried out as follows : local equilibrium condition between PBHs with temperature $T_{PBH}=\frac{1}{8\pi G M}$ and radius of horizon $r_{H}=2GM$ on the one hand and CMB with temperature $T_{\text{CMB}}=2.347\times 10^{-4}\text{eV}$ on the other hand  i.e. $T_{PBH} = T_{\text{CMB}}$, leads to $r_H\simeq 339  \, \text{eV}^{-1}$. Consequently the radius of PBH determines the cut-off wavelength i.e. $\lambda_c\sim r_H$ and thereby $k_c=\frac{2\pi}{\lambda_c}=\frac{2\pi}{r_H}=\frac{\pi}{GM}\simeq 1.85\times 10^{-2}\, \text{eV}$. Inserting $k_c$ in the expression for vacuum energy density of a massless scalar field we end up with
\be\label{eq:1}
\rho_v=\frac{1 }{16\pi^2}k_c^4 \simeq 74.6\times 10^{-11} \text{eV}^4
\ee
which is fairly close to the observed critical density $\rho_c\simeq 3.96\times 10^{-11}\text{eV}^4 $ \cite{Frieman:2008sn}. Mass of such  PBHs would be about $10^{26}$ gr which is still inside the mass window assigned to them through different observational constraints in scenarios with PBHs as dark matter candidates \cite{Capela:2013yf,Abramowicz:2008df,Seto:2007kj}. 
%%%%%%%%%%%%%%%%%%%%%%%%%%%%%%%%%%%%%%%%%%%%%%%%55
\section{The full scenario}
In the previous section, in our order of magnitude calculation, we implicitly inferred  $\rho_v \propto T_{PBH}^4$ as an estimate of the vacuum energy density but it should be noted that the relation used for the black hole temperature, $T_{PBH}=\frac{1}{8\pi G M}$, is only valid at infinity since it represents the temperature of a black hole of Mass $M$  as measured by an observer at infinity.  Since the equilibrium condition between CMB and the PBHs is employed at some point outside the black hole in the vicinity of its horizon, in a more complete scenario the following relation for the temperature of the black hole in an arbitrary distance  $r$ from its center  \cite{York:1986it}
\be\label{temperature}
T(r)=\frac{1}{8\pi G M}\left(1-\frac{2GM}{r}\right)^{-\frac{1}{2}}. 
\ee 
should be used. The above equation is a cubic equation for  $M$ in terms of $T$ and $r$ which means that the relation between $T$ and $M$  is not unique. Now assuming $r\geq0$ and $T\geq0$ and provided  $rT\geq\frac{\sqrt{27}}{8\pi}$, one can solve the above equation for $M$ to obtain the following two approximate solutions \cite{York:1986it}
\be\label{stablesolution}
M_1\simeq\frac{1}{2G}r\left(1-\frac{1}{(4\pi rT)^2}\right),
\ee
\be\label{unstablesolution}
M_2\simeq\frac{1}{8\pi G T}\left(1+\frac{1}{8\pi rT}\right).
\ee 
One can consider the system (CMB + PBH) as a canonical ensemble, assuming there exists a thermal bath of radius $r$ and temperature $T$ with which the black hole is in local thermal equilibrium. Such a system can be in a stable equilibrium state if its heat capacity is positive. The heat capacity is defined by $C\equiv\frac{\partial E}{\partial T}$, thus one needs to calculate $E$, the total thermodynamical energy of the system, which can be obtained by employing the Euclidean action. In fact $E$ is given by $E=-\frac{\partial \text{ln}\,Z}{\partial \beta}$, where $Z$ is the partition function of the theory and $\beta$ is the inverse temperature. In saddle point approximation, $\text{ln}Z\simeq- S_E$ where $S_E$ is the Euclidean action and so for the energy we have \cite{York:1986it}

\be\label{energy}   
E=\left(\frac{\partial S_E}{\partial \beta}\right)_r=r-r\left(1-\frac{2GM}{r}\right)^{-\frac{1}{2}},
\ee

\be\label{heat capacity} 
C_r=\left(\frac{\partial E}{\partial T}\right)_r=8\pi M^2\left(1-\frac{2GM}{r}\right)\left(\frac{3GM}{r}-1\right)^{-1}.
\ee
\vspace{12pt}
As we can see from the above equation $C_r$ would be positive if  $2GM<r<3GM$. This condition is the necessary but not the  sufficient condition to have a positive heat capacity. To find out which one of the solutions $M_1$ and $M_2$ leads to positive heat capacity we have
\begin{eqnarray}
C_r\bigg\rvert_{M=M_1}&=&\left( \frac{\partial E(M_1, r)}{\partial T}\right)_r=\left(\frac{\partial E(M_1, r)}{\partial M_1}\right)_r\left(\frac{\partial M_1}{\partial T}\right)_r\\\nonumber
&=&\frac{1}{\sqrt{1-\frac{2GM_1}{r}}}\frac{1}{(4\pi)^2GrT^3}>0
\end{eqnarray}
and
\begin{eqnarray}
C_r\bigg\rvert_{M=M_2}&=&\left( \frac{\partial E(M_2, r)}{\partial T}\right)_r=\left(\frac{\partial E(M_2, r)}{\partial M_2}\right)_r\left(\frac{\partial M_2}{\partial T}\right)_r\\\nonumber
&=&\frac{1}{\sqrt{1-\frac{2GM_2}{r}}}\left(-\frac{1}{8\pi GT^2}-\frac{2}{(8\pi)^2G rT^3}\right)<0.
\end{eqnarray}
So by the above argument it is only the $M_1$ solution which could be in stable equilibrium with the thermal bath enclosing the black hole. Therefore if we pick $M_1$ as  the black hole mass and assume that the surface of the spherical thermal bath (radius $r$) is located somewhere between $r=2GM_1$ and $r=3GM_1$, then the system (black hole and thermal bath) will be in stable local thermal equilibrium. Hereafter we drop the index of $M_1$ for convenience.\\
Now, through our scenario, we identify the  CMB as the above mentioned thermal bath. This identification  fixes the temperature of the black hole but not  the radius of the spherical  thermal bath, i.e knowing the BH temperature does not fix its mass. This is also obvious from  the expression (\ref{stablesolution})  in which by setting  $T=T_{\text{CMB}}$, the black hole  mass would still depend on $r$. Therefore for a black hole in thermal equilibrium with the CMB, there would be an infinite number of possible values for its mass. Setting $r=xGM$ in (\ref{stablesolution}), where $x\in(2, 3)$, we have
\be\label{modifiedmass}
M=\frac{1}{4\pi GT\sqrt{x^2-2x}}, 
\ee 
where $T$ is the  CMB temperature (Its  index is dropped for convenience and hereafter $T$ indicates the CMB temperature). The mass $M$ varies between two boundary values
\be\label{massspectrum}
M=\left\lbrace  \begin{array}{lr}
\infty &\quad x=2\\
\frac{1}{4\pi GT\sqrt{3}}\,\simeq2.9\times 10^{58} \text{eV}\, \,\text{for}\,\, T=2.7\,\, \text{K}\,\,&\qquad x=3
\end{array}\right.
\ee
As a result, in the equilibrium state with a given temperature, spectrum of the BH mass has a lower bound but no upper bound.
%%%%%%%%%%%%%%%%%%%%%%%%%%%%%%%%%%%%%%%%%%%%%%%%
\section{Dynamics of vacuum energy} 
In  the above mentioned scenario, the PBHs are supposed to be some sort of absorbers which  absorb vacuum fluctuations with wavelengths smaller than the radius of the PBHs. This absorption induces a natural IR cutoff ($\lambda_c=r_H$) on the wavelengths of vacuum fluctuations or equivalently a UV cutoff on their momenta. Hence we can write   
\be
\rho_v=\frac{k_c^4}{16\pi^2}=\frac{\pi^2}{\lambda_c^4}=\frac{\pi^2}{r_H^4}=\frac{\pi^2}{16 G^4}\frac{1}{M^4}
\ee 
If we interpret vacuum energy density as the dark energy density i.e. $\rho_\Lambda\equiv\rho_v$, we have 
\be\label{densityL}
\rho_\Lambda= \frac{\pi^2}{16 G^4}\frac{1}{M^4}.
\ee
Substituting (\ref{massspectrum}) in the above equation, spectrum of $M$ leads to the following spectrum for $\rho_\Lambda$,
\be\label{densityspectrum}
\rho_\Lambda=\left\lbrace  \begin{array}{lr}
0 &\quad x=2\\
144 \pi^6 T^4\simeq 42\times 10^{-11}\,\,\text{eV}^4\,\,\;\;\;\; \text{for}\,\, T=2.7 \,\, \text{K}&\quad x=3
\end{array}\right.
\ee
Since the PBHs in our model are taken as the  dark matter candidates, equation (\ref{densityL}) in principle represents a relationship/coupling between dark matter (PBHs of mass $M$) and dark energy (with density $\rho_\Lambda$) components of the Universe and this in turn could lead to a resolution of the cosmic coincidence problem \cite{Line}.
In order to determine the time evolution of $\rho_\Lambda$ we need to know the dynamics of $M$ in the course of the Universe's evolution. In (\ref{modifiedmass}), $T$ is the temperature of the CMB and  is given by $T=T_0 a^{-1}$ where $a$ is the scale factor in  the FLRW metric and the index $0$ indicates the present value of the CMB temperature. If we assume that $x$ in (\ref{modifiedmass}) is a constant, then  we will have $M\sim\frac{1}{T}$  implying that $\rho_\Lambda\sim T^4$ which is an unacceptable result. Therefore we restrict ourselves to the case in which $x$ is a function of the scale factor, i.e $x=x(a)$. Consequently, $M$ will be an arbitrary function of $a$. 
In order to determine  functionality of $M(a)$, we can use the energy conservation law 
\be\label{energyconservation}
\dot{\rho}+3H(\rho+p)=0,
\ee
where 
\be\label{densitypressure}
\rho =\rho_m+\rho_b+ \rho_\gamma+\rho_\Lambda,\qquad p= p_m +p_b+ p_\gamma + p_\Lambda,
\ee
and 
\be\label{pressure}
p_m=p_b=0,\quad p_\gamma=\tfrac{1}{3}\rho_\gamma,\quad p_\Lambda=w_\Lambda\rho_\Lambda,
\ee
in which $\rho_m=Mn$ is the dark matter density, $n$ is the number density of PBHs, $\rho_b=\rho_{0b}a^{-3}$ is the baryonic energy density, $\rho_\gamma=\frac{\pi^2}{15}T^4$ is the energy density of CMB and $w_\Lambda\simeq-1$. Now assuming that the total number of PBHs is conserved, we have
\be\label{numberdensity}
 \dot{n}+3Hn=0,
\ee
implying that $n=n_{0}a^{-3}$. Substituting  (\ref{densitypressure}), (\ref{pressure}) and (\ref{numberdensity} ) in Eq.(\ref{energyconservation}) leads to
\be\label{diffirentialequation1}
\dot{M}n+\dot{\rho}_\Lambda+3(1+w_\Lambda)H \rho_\Lambda=0.
\ee
Taking into account that $\rho_\Lambda$ is a function of $M$ and by defining $M'=\frac{dM}{da}$, (\ref{diffirentialequation1}) becomes
\be\label{diffirentialequation2}
M'\left(n+\frac{d\rho_\Lambda}{dM}\right)+\frac{3}{a}(1+w_\Lambda)\rho_\Lambda=0,
\ee
again using (\ref{numberdensity}) and (\ref{densityL}), we obtain
\be\label{diffirentialequation3}
M'\left(n_0 a^{-3}-\frac{\pi^2}{4G^4 M^5}\right)+\frac{3}{a}(1+w_\Lambda)\frac{1}{16G^4 M^4}=0.
\ee
This equation is very sensitive to the value of $w_\Lambda $ such that if it is chosen to be $-1$, the above differential equation will reduce to an algebraic equation, while a very small deviation from $-1$ leads to a very different behavior in comparison to the solution of that algebraic equation. Thus we consider the two different cases $w_\Lambda=-1$ and $w_\Lambda\neq-1$ separately.
%%%%%%%%%%%%%%%%%
\subsection{Case I: $w_\Lambda=-1$}
In this case, equation in (\ref{diffirentialequation3}) reduces to 
\be\label{algebraicequation}
M'=0\qquad \text{or}\qquad n_0a^{-3}-\frac{\pi^2}{4G^4M^5}=0
\ee
$M'=0$ leads to $M=$ constant, which is not an interesting case but the second equation gives
\be\label{mass1}
 M(a)=\left(\frac{\pi^2}{4G^4n_{0}}\right)^{\frac{1}{5}}a^{\frac{3}{5}}.
\ee
Using this solution in $\rho_m=Mn$  and  (\ref{densityL}), results in 
\be\label{matterdensity}
\rho_m=n_0^\frac{4}{5}\left(\frac{\pi^2}{4G^4}\right)^\frac{1}{5}a^{-\frac{12}{5}},
\ee
\be\label{Lambdadensity}
\rho_\Lambda=\frac{1}{4}n_0^\frac{4}{5}\left(\frac{\pi^2}{4G^4}\right)^\frac{1}{5}a^{-\frac{12}{5}}.
\ee
As pointed out earlier, in this scenario the  matter component couples to the dark energy component through equation  (\ref{densityL}) and this coupling gives a possible resolution to the coincidence problem. This issue  could be seen very clearly by inspection of   equations (\ref{matterdensity}) and (\ref{Lambdadensity}) from  which $\rho_\Lambda=\frac{1}{4}\rho_m$. 
%%%%%%%%%%%%%%%%%%%%%%%%%%%%%%5
\subsection{Case II: $w_\Lambda\neq-1$}
\begin{figure}
\includegraphics[scale=1]{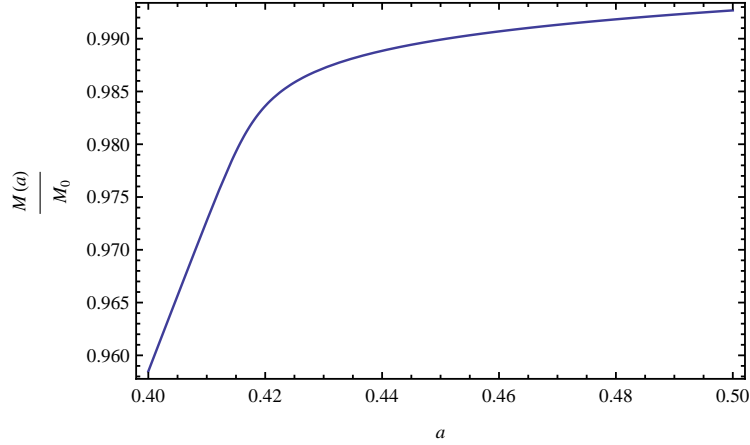}
\caption{$\frac{M(a)}{M_0}$ in terms of $a$  for $w_\Lambda=-0.99$. }\label{figmass}
\end{figure}
\begin{figure}
\includegraphics[scale=1.05]{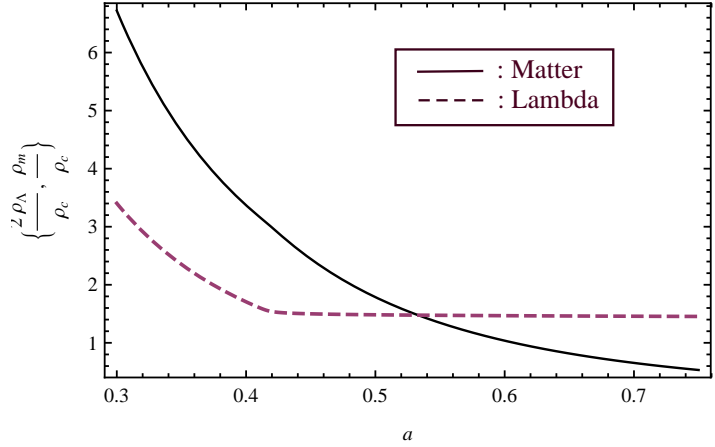}
\caption{$2\rho_\Lambda$(dashed curve)  and $\rho_m$(solid curve) in terms of $a$.}\label{figdensity}
\end{figure}

In this case, differential equation in (\ref{diffirentialequation3}) should be solved numerically for which we need one boundary condition and also the numerical value for the $n_0$. Using the present values of $\rho_{0m}=0.22\rho_c$ and $\rho_{0\Lambda}=0.72\rho_c$ and substituting them into  (\ref{densityL}) and also in $\rho_{0m}=M_0n_0$, we end up with,
\be\label{numericalvalue}
M_0=5.7\times10^{58} \text{eV},\qquad
 n_0=1.5\times10^{-70}{\text{eV}}^3.
\ee   
As an example we have solved the differential equation  (\ref{diffirentialequation3}) for $w_\Lambda=-0.99$, the result of which is plotted in Fig. \ref{figmass}.  Having found  the functionality of $M(a)$, the behavior of matter density and dark energy density could also be obtained. The results of which are depicted in Fig. \ref{figdensity}. In  a Universe dominated by matter and dark energy (with an equation of state with $w\simeq-1$), there is a transition point from  deceleration to acceleration at $\rho_m\simeq 2\rho_\Lambda$. To find this transition point in our scenario, in Fig. \ref{figdensity} we plotted  $2\rho_\Lambda$ against $\rho_m$ which shows that this transition can happen around $a\simeq0.54 $ which is close to the  value predicted by the  $\Lambda$CDM model \cite{Komatsu:2010fb}  

\begin{figure}
\includegraphics[scale=1]{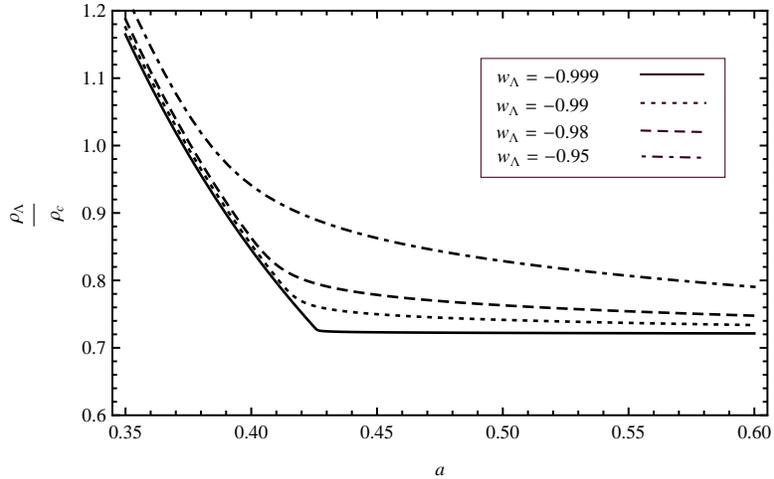}
\caption{Evolution of dark energy density for several values of $w_\Lambda$}.\label{figdensity2}
\end{figure}

As  seen from Fig. \ref{figdensity}, for $a\lesssim 0.42$, behavior of $\rho_\Lambda$  is the same as that of matter density as it  happenes in the $w_\Lambda=-1$ case. For $a\gtrsim 0.42$, $\rho_\Lambda$  is nearly constant and such a behavior is related to the  deviation of $w_\Lambda$ from $-1$ and is totally different from that of $\rho_\Lambda$  in  the $w_\Lambda=-1$ case. It is the slow variation of $\rho_\Lambda$ for  $a\gtrsim 0.42$ that allows transition from  deceleration to acceleration. Now the question may arise how  a small deviation from $w_\Lambda=-1$ is possible?. In fact, this deviation can happen due to the introduction of the interaction terms in the original Lagrangian of the scalar field and  also  due to the effects of time-dependent curved  background. Fig. \ref{figdensity2} shows the evolution of $\rho_\Lambda$ for several values of $w_\Lambda$, all of which approaching the same value at the present epoch. Also, as could be seen, the more $w_\Lambda$ is closer to $-1$, the more $\rho_\Lambda$ gets close to a constant value after $a\approx 0.42$ but it switches smoothly to a nearly constant evolution when deviation from $-1$ is increased.  

%%%%%%%%%%%%%%%%%%%%%%%%%%%%%%%%%%%%%%5
\section{Concluding remarks}
In the present article, we have introduced a scenario in order to obtain a finite zero-point energy density for an effective scalar field as the dark energy density. The energy density is found to be of the same order as the  measured value through different cosmological observations. We showed that  PBHs with an appropriate absorption  cross section can restrict the modes of the vacuum by absorbing those modes whose wavelengths are smaller than the radius of PBHs.  The appropriate absorption cross section was obtained by imposing the local equilibrium condition between CMB and PBHs. For establishing such an equilibrium we have employed  York's approach \cite{York:1986it} which  shows that in contrast to a black hole in an empty space which can not be in stable equilibrium, the same black hole  can be in stable equilibrium if it is surrounded  by a radiation (in our scenario CMB) as a spherical thermal bath  covering the black hole to a given radius from its center\cite{York:1984wp,Hochberg:1993ps}. Absorption of vacuum mode  by the PBH  increases
PBH's mass and this in turn gives rise to a decrease of the vacuum energy density of a scalar field which was considered as the dark energy candidate.  In this way we have introduced a variable dark energy term which is coupled to the dark matter component and through this coupling the same scenarion was also capable of resolving  the so called cosmological coincidence problem.

%%%%%%%%%%%%%%%%%%%%%%%%%%%%%%%%%%%5
\begin{acknowledgements}
The authors would like to thank M. M. Sheikh-Jabbari for useful discussions. S. Z. also thanks A. Davody for helpful  discussions and  M. Golshani for encouragement and support. M. N-Z thanks University of Tehran's office of research affairs for partial financial support.
\end{acknowledgements}


\begin{thebibliography}{99}

\bibitem{Weinberg:1989cp} 
  S.~Weinberg,
  %``The Cosmological Constant Problem,''
  Rev.\ Mod.\ Phys.\  {\bf 61}, 1 (1989).
  %%CITATION = RMPHA,61,1;%%
\bibitem{Freese:1987dd} 
  K.~Freese, F.~C.~Adams, J.~A.~Frieman and E.~Mottola,
  %``Cosmology with Decaying Vacuum Energy,''
  Nucl.\ Phys.\ B {\bf 287}, 797 (1987).
  %%CITATION = NUPHA,B287,797;%%
\bibitem{Overduin:1998zv} 
  J.~M.~Overduin and F.~I.~Cooperstock,
  %``Evolution of the scale factor with a variable cosmological term,''
  Phys.\ Rev.\ D {\bf 58}, 043506 (1998)
  [astro-ph/9805260].
  %%CITATION = ASTRO-PH/9805260;%%
\bibitem{Opher:2004}
  R. Opher and and A. Pelinson, Phys. Rev. D., 70, 063529 (2004) ; R. Opher and and A. Pelinson, 
  Mon. Not. Roy. Astron.Soc. 362 (2005) 167-170.
\bibitem{Frampton:2010sw} 
  P.~H.~Frampton, M.~Kawasaki, F.~Takahashi and T.~T.~Yanagida,
  %``Primordial Black Holes as All Dark Matter,''
  JCAP {\bf 1004}, 023 (2010) [arXiv:1001.2308 [hep-ph]].
  %%CITATION = ARXIV:1001.2308;%%
  %27 citations counted in INSPIRE as of 01 Sep 2013
\bibitem{Blais}
  D. Blais, C. Kiefer and D. Polarski, Phys. Lett. B {\bf 535}, 11 (2002).
\bibitem{Khlop}
  M. Y. Khlopov, Res. Astron. Astrophys. {\bf 10}, 495 (2010).
\bibitem{Martin} 
  J. Martin, Comptes Rendus Physique
  , Volume 13, Issues 6–7, 2012, 566 [arXiv:1205.3365 [astro-ph]].
\bibitem{Paddy} 
  T.~Padmanabhan,
  %``Vacuum fluctuations of energy density can lead to the observed cosmological constant,''
  Class.\ Quant.\ Grav.\  {\bf 22}, L107 (2005) [hep-th/0406060].
  \bibitem{Frieman:2008sn} 
  J.~Frieman, M.~Turner and D.~Huterer,
  %``Dark Energy and the Accelerating Universe,''
  Ann.\ Rev.\ Astron.\ Astrophys.\  {\bf 46}, 385 (2008) [arXiv:0803.0982].
  %%CITATION = ARXIV:0803.0982;%%  
\bibitem{Seto:2007kj} 
  N.~Seto and A.~Cooray,
  %``Searching for primordial black hole dark matter with pulsar timing arrays,''
  Astrophys.\ J.\  {\bf 659}, L33 (2007) [astro-ph/0702586].
  %%CITATION = ASTRO-PH/0702586;%% 
  \bibitem{Abramowicz:2008df} 
  M.~A.~Abramowicz, J.~K.~Becker, P.~L.~Biermann, A.~Garzilli, F.~Johansson and L.~Qian,
  %``No observational constraints from hypothetical collisions of hypothetical dark halo primordial black holes with galactic objects,''
  Astrophys.\ J.\  {\bf 705}, 659 (2009)
  [arXiv:0810.3140 [astro-ph]].
  %%CITATION = ARXIV:0810.3140;%%   
 \bibitem{Capela:2013yf} 
  F.~Capela, M.~Pshirkov and P.~Tinyakov,
  %``Constraints on primordial black holes as dark matter candidates from capture by neutron stars,''
  Phys.\ Rev.\ D {\bf 87}, 123524 (2013)
  [arXiv:1301.4984 [astro-ph]].
  %%CITATION = ARXIV:1301.4984;%%
  %1 citations counted in INSPIRE as of 01 Sep 2013 


 \bibitem{York:1986it} 
  J.~W.~York, Jr.,
  %``Black hole thermodynamics and the Euclidean Einstein action,''
  Phys.\ Rev.\ D {\bf 33}, 2092 (1986).
  %%CITATION = PHRVA,D33,2092;%% 
  \bibitem{Line}
   C. Egan and C. H. Lineweaver,  Phys. Rev. D {\bf78}, 083528 (2008). 
\bibitem{Komatsu:2010fb} 
  E.~Komatsu {\it et al.}  [WMAP Collaboration],
  %``Seven-Year Wilkinson Microwave Anisotropy Probe (WMAP) Observations: Cosmological Interpretation,''
  Astrophys.\ J.\ Suppl.\  {\bf 192}, 18 (2011) [arXiv:1001.4538 [astro-ph.CO]].
  %%CITATION = ARXIV:1001.4538;%% 

\bibitem{York:1984wp} 
  J.~W.~York, Jr.,
  %``Black Hole In Thermal Equilibrium With A Scalar Field: The Back Reaction,''
  Phys.\ Rev.\ D {\bf 31}, 775 (1985).
  %%CITATION = PHRVA,D31,775;%%
  %138 citations counted in INSPIRE as of 01 Sep 2013   
\bibitem{Hochberg:1993ps} 
  D.~Hochberg, T.~W.~Kephart and J.~W.~York, Jr.,
  %``Effective potential of a black hole in thermal equilibrium with quantum fields,''
  Phys.\ Rev.\ D {\bf 49}, 5257 (1994) [gr-qc/9307037].
  %%CITATION = GR-QC/9307037;%%
  %11 citations counted in INSPIRE as of 01 Sep 2013
\end{thebibliography}
\end{document}